\documentclass[twocolumn,aps,prb,amsmath,amssymb,preprintnumbers,superscriptaddress,longbibliography]{revtex4-2}
\usepackage[utf8]{inputenc}
\usepackage[english]{babel}
\usepackage{physics}
\usepackage[normalem]{ulem}
\usepackage{braket}
\usepackage{bbold}
\usepackage{amsmath, amsfonts, amssymb}
\usepackage{graphicx}
\usepackage{float}
\usepackage{lipsum}
\usepackage[colorlinks=true,linkcolor=blue,citecolor=blue,allcolors=blue]{hyperref}
\usepackage[caption=false]{subfig}
\usepackage{mathtools}
\usepackage{tikz}
\usepackage{natbib}
\usepackage{soul}

\newcommand{\bx}{\mathbf{x}}
\newcommand{\by}{\mathbf{y}}
\newcommand{\bs}{\mathbf{s}}
\newcommand{\ones}{\mathbb{1}}

\begin{document}
    	\newcommand{\titleinfo}{Learning Probability Distributions for Black-Box Optimization with Projected Entangled Pair States}
	\title{\titleinfo}

    \author{Aditya Dubey}
    \email{a.dubey@fz-juelich.de}
\affiliation{%
 Forschungszentrum J\"ulich  GmbH, Institute of Quantum Control, Peter Gr\"unberg Institut (PGI-8), 52425 J\"ulich, Germany
}
\affiliation{Institute for Theoretical Physics, University of Cologne, D-50937 K\"oln, Germany}

    \author{Zeki Zeybek}
    \email{zeki.zeybek@uni-hamburg.de}
	\affiliation{Zentrum f{\"u}r Optische Quantentechnologien, Universit{\"a}t Hamburg, Luruper Chaussee 149, 22761 Hamburg, Germany}

	\begin{abstract}
     Optimization algorithms are used virtually everywhere in science and engineering. They can be roughly categorized into gradient-based and gradient-free algorithms. Gradient-based algorithms are efficient when reliable derivative information is available. However, derivatives may be costly, noisy, or unavailable, for example when objective evaluations rely on measurements or the landscape is rugged. Gradient-free algorithms are useful in that setting, but representing and updating an effective search strategy becomes challenging in high-dimensional search spaces. Here, we consider a gradient-free probabilistic optimization framework employing projected entangled pair states (PEPS), which we call PEPS-Opt. In this approach, a probability distribution parameterized by a PEPS is iteratively refined to favor configurations with better objective values. We employ a DMRG-like local sweep algorithm that exploits the multilinear structure of the PEPS to update its tensors. We extend direct sampling to 1-norm parameterized PEPS and introduce quantics PEPS with Gray-code encoding for continuous optimization variables. The method is benchmarked on continuous functions, Ising spin glasses, and parameter optimization of the quantum approximate optimization algorithm (QAOA), using both exact and shot-noise-limited objective evaluations. Across these problems, PEPS-Opt achieves competitive performance with established gradient-free optimizers, proving particularly effective on rugged landscapes and for deeper QAOA circuits, while remaining robust under finite-shot measurement noise.

	\end{abstract}
	\maketitle

\section{Introduction} 
Optimization problems arise throughout science and engineering
and can become particularly challenging in high-dimensional
settings with nonconvex objective landscapes. For example,
discrete optimization problems can exhibit rugged energy
landscapes arising from frustrated interactions
~\cite{Edwards_SpinGlass,SpinGlass_EnergyLandscape}.
Similarly, optimization in variational quantum algorithms can
be challenging due to noisy objective evaluations that arise
from finite measurement statistics and hardware imperfections
~\cite{VQA_ShotNoise,Cerezo_VQA}.
In this sense, robust global optimization methods are especially
important when gradient information is unavailable or unreliable ~\cite{Kochenderfer_Algorithms_for_Optimization,Larson_DerivativeFree}.

Gradient-free methods look for a search strategy only from
objective evaluations and can include local, population-based,
and probabilistic approaches
~\cite{Kochenderfer_Algorithms_for_Optimization,Conn_DerivativeFree,Hansen_CMAES}.
In this work, we are interested in methods that maintain a
parameterized search distribution over candidate solutions and
iteratively update it using the performance of sampled
configurations~\cite{Larranaga_EDA,Rubinstein_CrossEntropy}.
In this way, the search distribution is progressively biased
toward configurations associated with lower objective values
~\cite{DeBoer_CrossEntropy}.

A main design objective in such probabilistic approaches is to determine how to represent, learn, and update the search distribution efficiently. This can be challenging when the problem at hand admits a high-dimensional search space. In this setting, tensor networks can be used as the adaptive search distribution over discrete candidate configurations. Tensor networks provide efficient parameterization for
functions~\cite{Oseledets_TT} and probability distributions
admitting low-rank structure~\cite{dolgov_sample} by
decomposing high-order tensors into networks of lower-order
tensors connected through virtual bonds~\cite{Cirac_TN_Review}. Originally developed in quantum many-body physics
\cite{White_DMRG,Verstaete_PEPSAlgo,AD_MCDTH_TTN,ZZ_DMRG_MB_RYDII},
they have since found many applications in machine learning
\cite{Stoudenmire_TN_ML,Han_MPS_Generative,Cheng_PEPS_Supervised,Vieijra_PEPS_Generative},
numerical analysis
\cite{Oseledets_TT,Holtz_TT_ALS,Khoromskij_QTT2011,dolgov_sample},
and quantum computing
\cite{Vidal_QuantumSimulation,Markov_Shi_QCirc,Gray_TN_Contraction,AD_DMRG_QCirc}.
However, existing tensor network optimization methods have predominantly relied on tensor train representations \cite{Alcazar_GenerativeOpt, Piqueres_symm, Piqueres_constrain, Piqueres_EDA, PROTES, ZZ_TTEDA}. Moving beyond this setting may be useful for optimization problems that naturally lend themselves to two-dimensional or higher-dimensional tensor network representations. At the same time, this raises fundamental computational problems associated with contraction, sampling, and optimization of higher-dimensional tensor networks.

\begin{figure*}[t]
    \centering
    \includegraphics[width=1\linewidth]{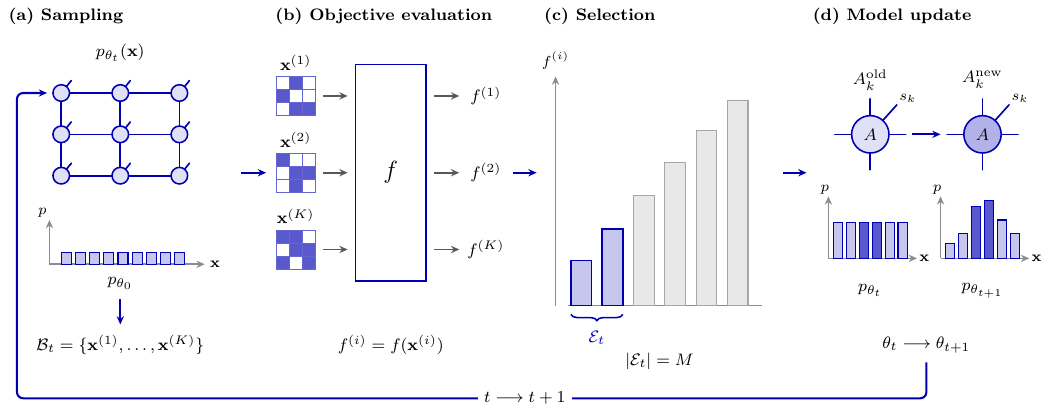}
    \caption{Probabilistic optimization with a PEPS search distribution.
(a) At iteration $t$, sample $K$ candidate configurations to form the set
$\mathcal B_t=\{\bx^{(i)}\}_{i=1}^{K}$ from $p_{\theta_t}(\bx)$.
The histogram illustrates the uniform initial distribution $p_{\theta_0}$.
(b) Evaluate the black-box objective to obtain
$f^{(i)}=f(\bx^{(i)})$.
(c) Rank the candidates by their objective values and retain the $M$
lowest-cost configurations as the elite set $\mathcal E_t$.
(d) Update the PEPS tensors using the elites to obtain $\theta_{t+1}$. Updating the tensors
reshapes the search distribution to favor these configurations. Repeat the loop until the budget for the objective evaluation is exhausted and
return the best configuration encountered.}
\label{fig:general_framework}
\end{figure*}

In this work, we introduce a probabilistic optimization
framework for projected entangled pair states (PEPS),
which we refer to as PEPS-Opt, and address the associated
algorithmic challenges
~\cite{Schuch_PEPS_Complexity,Lubasch_PEPSContactions}.
A PEPS arranges local tensors on a two-dimensional lattice,
with each bulk tensor connected to four nearest neighbors
through virtual bonds in addition to its physical index
~\cite{Verstaete_PEPSAlgo,Verstraete_PEPS_AreaLaw}. We use this structure to parameterize a normalized probability distribution over discrete candidate configurations. Its two-dimensional connectivity is particularly appealing when the optimization variables exhibit a corresponding interaction structure, since variables coupled through the objective can be assigned to neighboring PEPS sites. Accordingly, our aim is to investigate how PEPS can be used as adaptive search distributions for discrete black-box optimization and to characterize their practical behavior, strengths, and limitations. To this end, we formulate a non-negative PEPS as a search distribution and extend the direct sampling algorithm to generate candidate solutions. We introduce Gray-code encoding for quantics~\cite{Khoromskij_QTT} representations of continuous objective functions and adopt a local sweeping algorithm for optimizing the PEPS tensors~\cite{lee_learning,hood2026nearuniversal}. We benchmark the proposed framework on a range of problems, including multimodal test functions with nontrivial landscapes, Ising spin glasses, and parameter optimization for the quantum approximate optimization algorithm (QAOA)~\cite{Farhi_QAOA}. We find that PEPS-Opt is most
competitive on rugged landscapes with many separated
local minima, such as multimodal functions and spin
glass problems, and that it remains robust when QAOA objectives are estimated from finite measurement shots. Our results indicate that the proposed approach provides a useful framework for gradient-free optimization across these different problem classes.

The paper is organized as follows. In Section \ref{seq:theory}, we introduce the optimization framework and detail the PEPS representation of the search distribution, sampling from 1-norm parameterized PEPS as well as update and optimization methods. We also introduce the optimization variable encoding strategies. Section \ref{seq:results} presents benchmark results across various optimization problems where we compare our method to standard gradient-free optimizers. Finally, Section \ref{seq:conc} concludes with a summary and outlook on future directions.

\section{Probabilistic Optimization with PEPS}
\label{seq:theory}

We consider a discrete black-box optimization problem of the form
\begin{equation}
    \mathbf{x}^{\star}
    =
    \arg\min_{\mathbf{x}\in\mathcal{X}} f(\mathbf{x}),
\end{equation}
where
\(
\mathbf{x}=(x_1,\ldots,x_N)
\)
denotes a configuration of discrete optimization variables and
\(f:\mathcal{X}\rightarrow\mathbb{R}\) is an \textit{objective function} that can be evaluated point-wise but whose derivatives with respect to the discrete variables are unavailable. For example, in quantum control, evaluating \(f(\mathbf{x})\) amounts to propagating the system dynamics under the control parameters specified by \(\mathbf{x}\) and computing a task-specific cost, such as the state or gate infidelity. In variational quantum algorithms such as QAOA, evaluating the objective requires executing the parameterized quantum circuit for a given set of variational angles.

Rather than constructing an approximation of the objective function \(f(\mathbf{x})\) itself,
we introduce a parameterized probability distribution
\(p_{\theta}(\mathbf{x})\) over the search space and adapt this distribution during the optimization. Candidate configurations are then sampled from the current distribution, evaluated with the black-box objective \(f(\mathbf{x})\), and a subset
of the best-performing configurations is kept. The parameters
\(\theta\) of the search distribution are then subsequently updated by increasing the probability assigned to these selected configurations. Repeating this procedure gradually reshapes the search distribution using information from previous \(f(\mathbf{x})\) evaluations. In this sense, our approach falls within the broader family of probabilistic optimization methods that maintain and iteratively adapt a parameterized search distribution, which includes the cross-entropy method and natural evolution strategies \cite{rubinstein2004crossentropy,wierstra2014natural}.

Tensor networks have recently been employed to parameterize adaptive search distributions for discrete and combinatorial optimization. In these approaches, candidate configurations are sampled from a tensor network probability model, and the tensors are iteratively updated using high-performing samples to shift probability mass toward promising regions of the search space \cite{Alcazar_GenerativeOpt, Piqueres_symm, Piqueres_constrain, Piqueres_EDA,PROTES,ZZ_TTEDA}. Our approach follows the same general probabilistic optimization framework, but replaces the tensor-train search distribution by a projected entangled-pair state (PEPS). In this way, the resulting model parameterizes the probability distribution over discrete candidate configurations with a two-dimensional tensor network. This can be particularly suitable when the optimization variables can be arranged on a two-dimensional lattice such that the lattice connectivity reflects the dependency or interaction structure of the problem. We outline the optimization loop depicted in Fig.~\ref{fig:general_framework} as follows:
\begin{enumerate}
    \item \textbf{Initialization:}
    Initialize a PEPS with parameters \(\theta_0\) defining a normalized search distribution \(p_{\theta_0}(\mathbf{x})\). The initial distribution is chosen sufficiently broad to provide coverage of the admissible search space.
    \item \textbf{Sampling:} During iteration \(t\), 
    generate \(K\) candidate configurations
    \(
    \mathcal{B}_t
    =
    \{\mathbf{x}^{(i)}\}_{i=1}^{K}
    \)
    from the current PEPS search distribution
    \(p_{\theta_t}(\mathbf{x})\).
    \item \textbf{Objective evaluation:}
    Evaluate the black-box objective for every sampled configuration,
    \begin{equation}
        f^{(i)}
        =
        f\!\left(\mathbf{x}^{(i)}\right),
        \qquad i=1,\ldots,K.
    \end{equation}
    \item \textbf{Selection:}
    Rank the sampled configurations according to their objective values
    and retain the \(M\) best-performing configurations as the elite set \(\mathcal{E}_t\).
    \item \textbf{Model update:}
    Update the PEPS parameters by increasing the probability given to selected elite configurations. The resulting parameters \(\theta_{t+1}\) then define the search distribution used in the next iteration.
    \item \textbf{Iteration:}
    Repeat the sampling, evaluation, selection, and update
    steps until the predetermined objective evaluation budget is exhausted.
\end{enumerate}

\subsection{PEPS as Search Distribution}
\label{subsec:peps_as_search_distribution}
\begin{figure*}
    \centering
    \includegraphics[width=1\linewidth]{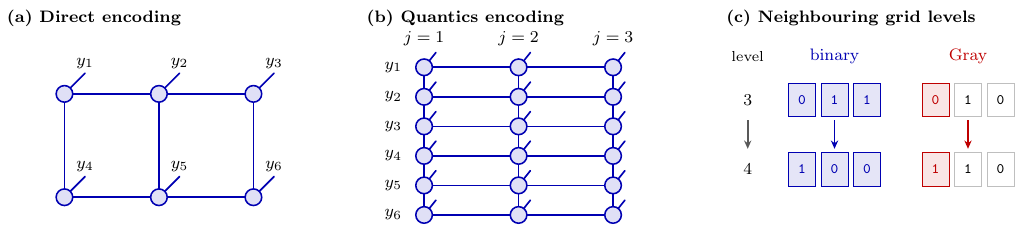}
    \caption{Encoding schemes. Blue circles represent the site tensors;
links between circles are virtual bonds of dimension $D$, and outward
legs carry physical indices.
(a) Direct encoding assigns each of $V$ variables to one site of a PEPS.
For a continuous variable discretized to $2^R$ levels, the physical
dimension is $d=2^R$; six variables are shown on a rectangular lattice.
(b) Quantics encoding resolves the same variables on a $V\times R$ grid
with one bit per site and physical dimension $d=2$. Rows label variables
and columns their binary digits, ordered from most to least significant
from left to right; $V=6$ and $R=3$ are shown. Horizontal bonds connect
digits of the same variable, while vertical bonds couple different variables.
(c) Consecutive levels $3$ and $4$ on the three-bit grid correspond to
$y=3/8$ and $4/8$. Their binary codewords differ in three bits,
whereas their reflected Gray codewords differ in one bit. Colored boxes
mark the bits that change. Both codes represent the same grid points.}
    \label{fig:encoding}
\end{figure*}

Projected entangled pair states (PEPS) provide a low-rank representation of high-dimensional tensors as a two-dimensional network of small tensors. Here, we employ PEPS to parameterize the probability distribution over a high-dimensional search space
of variables. We consider an $L_x\times L_y$ rectangular lattice with
$N=L_xL_y$ sites labeled by $k=(i,j)$. The physical index
$s_k\in\{0,\ldots,d-1\}$ is a free tensor index that runs through available local states.  Here, it labels the values that each optimization variable can take. A single site tensor of PEPS is written as $A_k$ with entries
$A^{s_k}_{k,\boldsymbol\alpha_k}$. The multi-index
$\boldsymbol\alpha_k=(\alpha_l,\alpha_r,\alpha_u,\alpha_d)$ collects
its virtual indices, each of which is dimension $D$:

\begin{equation}
  A^{s_k}_{k,\boldsymbol\alpha_k}
  =\vcenter{\hbox{\includegraphics[width=.36\linewidth]{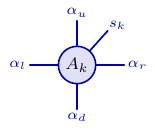}}}.
  \label{eq:site_tensor_diagram}
\end{equation}
Then the network weight is expressed as
\begin{equation}
  \begin{aligned}
  W_\theta(\textbf{s})
    &=\operatorname{tcon}\!\left[\prod_k A_k\right]\\
    &=\sum_{\{\alpha\}}\prod_k A^{s_k}_{k,\boldsymbol\alpha_k},
  \end{aligned}
  \label{eq:peps}
\end{equation}
where $\theta=\{A_k\}_{k=1}^N$ stands for the collection of all variational parameters and $\operatorname{tcon}$ contracts all shared virtual bonds.
The corresponding tensor network representation is
\begin{equation}
  W_\theta(\bs)
  =\vcenter{\hbox{\includegraphics[width=.48\linewidth]{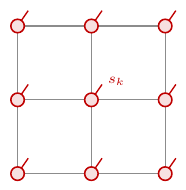}}}.
  \label{eq:weight_tensor_diagram}
\end{equation}
For a specific configuration, $W_\theta(\bx)$ is obtained by fixing
$s_k=x_k$ at every site,
\begin{equation}
  W_\theta(\bx)
  =\vcenter{\hbox{\includegraphics[width=.48\linewidth]{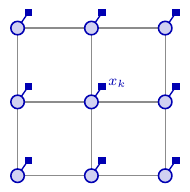}}}.
  \label{eq:configuration_weight_diagram}
\end{equation}
The square physical-leg endpoints denote projection onto the specified
values. With every index contracted, the network gives a scalar weight.

To represent the PEPS ansatz as a valid probability distribution, two different parametrizations can be considered. The first is the \emph{Born-machine} formulation, which mirrors the standard use of PEPS in quantum many-body physics where \(W_\theta(\bx)\) represents a wavefunction amplitude. The tensors may therefore have signed or complex entries, and the probabilities are obtained through the Born rule.

In this formulation, computing the normalization and marginal probabilities then involves contracting a double-layer tensor network with virtual bond dimension \(D^2\). In this work, we instead adopt a \emph{1-norm} parametrization in which all entries of the site tensors \(A_k\) are non-negative. Consequently, normalization and marginalization can be performed by contracting a single-layer network with bond dimension \(D\) rather than the $D^2$ network arising in the Born-machine formulation. The probability distribution is then defined directly as
\begin{equation}
  p_\theta(\bx)
  =
  \frac{W_\theta(\bx)}{Z_\theta},
  \qquad
  Z_\theta
  =
  \sum_{\bs}W_\theta(\bs).
  \label{eq:onenorm}
\end{equation}
Here, $W_\theta(\bx)$ is an unnormalized probability weight.
$Z_\theta$ is obtained by summing each physical index, which is
implemented by contracting every physical leg with the all-ones vector
$\ones_d$,
\begin{equation}
  Z_\theta
  =\vcenter{\hbox{\includegraphics[width=.48\linewidth]{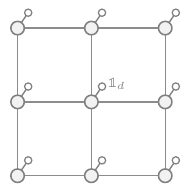}}}.
  \label{eq:normalization_diagram}
\end{equation}
The circular endpoints denote these sums. The ansatz contains \(O(NdD^4)\) tensor entries. With the probability distribution defined, we next describe how the variables of the objective function are encoded in the physical indices of the PEPS. We consider two encoding schemes.

\paragraph{Direct encoding.}
Each variable is assigned to a single site, as illustrated in Fig.~\ref{fig:encoding}(a). For a discrete
variable, $x_k\in\{0,\ldots,d-1\}$ labels its possible values; for a
continuous variable, it labels the points of a chosen discretization.
Thus, each configuration $\mathbf{x}$ specifies one point in
the search space. This encoding is particularly suitable for problems
whose variables and interactions are naturally arranged on a
lattice, such as spin-glass models. Its main limitation is that the required physical dimension grows with the number of values used to represent each variable which can become costly for finely discretized continuous variables.

\paragraph{Quantics encoding.}
Quantics encoding avoids the growth of the local physical dimension by representing each variable through a sequence of binary indices. The more significant bits determine the value on a coarse scale, while the less significant bits progressively refine it. This binary representation therefore introduces a natural coarse-to-fine structure. Quantics tensor trains have successfully exploited this structure for compact function representation and numerical problems such as partial differential equations~\cite{Khoromskij_QTT,Khoromskij_QTT2011}. A quantics PEPS extends the same idea to two dimensions by arranging the binary indices on a lattice: rows label variables, while columns label their binary digits [Fig.~\ref{fig:encoding}(b)]. 

Consider an objective function $f(y_1,\ldots,y_{V})$, with
$y_i\in[0,1)$. Each variable is discretized on a uniform grid of
$2^{R}$ points with spacing $2^{-R}$. A grid point is specified
by
\begin{equation}
  y_i(\mathbf{x})
  =
  \sum_{j=1}^{R} x_{(i,j)}\,2^{-j},
  \qquad
  x_{(i,j)}\in\{0,1\}.
  \label{eq:quantics_encoding}
\end{equation}
The resulting search space is therefore encoded by
$N=V R$  binary indices on a $V \times R$ PEPS 
with physical dimension $d=2$. Here, $V$ is the number of variables and $R$ the number of bits
per variable. Site $(i,j)$ carries the $j$-th binary digit of variable $y_i$.
Horizontal bonds connect successive digits of the same variable,
while vertical bonds couple the encodings of different variables.

The standard binary encoding in Eq.~\eqref{eq:quantics_encoding} provides a natural separation of scales, but it does not preserve locality between neighboring grid levels. Adjacent levels can differ in several bits (e.g.\ $0111\to1000$), creating a ``Hamming cliff''. As a result, values that are close in the original variable may correspond to configurations that differ across several physical indices. We instead consider Gray encoding, in which consecutive levels differ by exactly one bit, as illustrated in Fig.~\ref{fig:encoding}(c). The physical indices $x_{(i,j)}$ then store Gray-code bits, and Eq.~\eqref{eq:quantics_encoding} is replaced by
\begin{equation}
  b_{(i,j)}=\bigoplus_{m=1}^{j}x_{(i,m)},
  \qquad
  y_i(\mathbf{x})=\sum_{j=1}^{R}b_{(i,j)}\,2^{-j},
\end{equation}
where $\oplus$ denotes XOR. Both encodings represent the same $2^R$ discretization levels per variable and retain a hierarchical refinement of the numerical domain. Gray encoding additionally ensures that consecutive levels differ at only one physical index.

\subsection{Sampling from PEPS}
\label{subsec:Sampling}

\begin{figure*}
    \centering
    \includegraphics[width=1\linewidth]{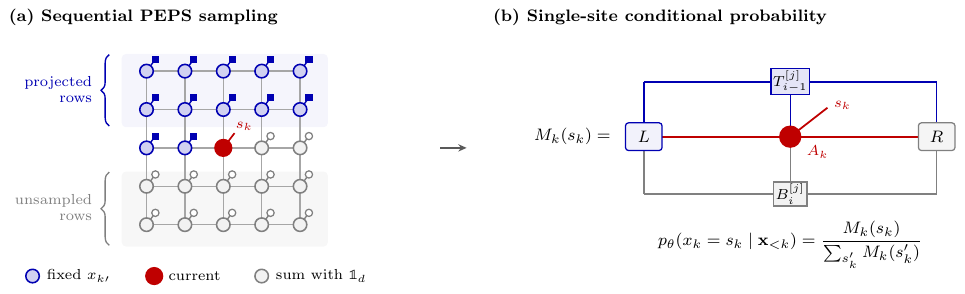}
    \caption{Single-layer  sampling of a 1-norm PEPS.
(a) Previously sampled sites are projected on their sampled values
(blue; square physical-leg), the current site is left open
(red), and later sites are summed over with $\ones_d$
(gray; circular physical-leg). (b) The current tensor $A_k$ is contracted with the $j$th tensors of the
top and bottom boundary MPS, $T_{i-1}^{[j]}$ and
$B_i^{[j]}$, and the row environments $L$ and $R$.
$L$ contains the already sampled sites to the left, with their physical
indices fixed; $R$ contains the unsampled sites to the right, with their
physical indices summed. Both include the corresponding top and bottom
boundary tensors.
Their contraction with the current column gives the length-$d$ vector
$M_k$, whose normalized entries determine the drawing probability of $x_k$.}
    \label{fig:sampling}
\end{figure*}

Efficient sampling from the search distribution is essential for generating candidate configurations during optimization. To this end, we adapt the direct-sampling method developed for PEPS in the Born-rule formulation~\cite{PEPS_Direct_Sampling} to the non-negative 1-norm distribution introduced in Eq.~\ref{eq:onenorm}. Unlike methods using Markov chains~\cite{Liu_GradOptGroundState}, direct sampling generates configurations independently. This helps avoid autocorrelations and the need for specific update schemes for each problem. The independence of the generated configurations also allows sampling to be embarrassingly parallel.
The central idea is to construct configurations sequentially: sampling each physical index from its conditional probability given the previously sampled indices. We obtain these conditional probabilities by marginalizing over the remaining physical indices. In this way, we reduce the sampling problem to a sequence of tensor 
network contractions as shown in Fig.~\ref{fig:sampling}. We describe the sampling procedure row by row. Let
\begin{equation}
    \bx_{[i]} =
    (x_{(i,1)},x_{(i,2)},\ldots,x_{(i,L_y)})
\end{equation}
denote the configuration of row $i$, and let
$\bx_{[<i]}=(\bx_{[1]},\ldots,\bx_{[i-1]})$
collect the previously sampled rows. The full configuration
is written as
$\bx=(\bx_{[1]},\ldots,\bx_{[L_x]})$.
Using the chain rule of probability, the joint distribution
can be decomposed exactly as
\begin{equation}
    p_\theta(\bx)
    =
    \prod_{i=1}^{L_x}
    p_\theta(\bx_{[i]}\mid\bx_{[<i]}),
    \label{eq:row_chain}
\end{equation}
where the first factor is simply $p_\theta(\bx_{[1]})$. Evaluating each conditional requires combining the
contribution from the previously sampled rows with the
marginalization over all subsequent rows.

For a fixed row configuration $\bx_{[i]}$, we define
the single-layer row transfer operator
\begin{equation}
    \begin{aligned}
    &[C_i(\bx_{[i]})]_{\mathbf d,\mathbf u}\\
    &=
    \sum_{h_1,\ldots,h_{L_y-1}}
    \prod_{j=1}^{L_y}
    A^{x_{(i,j)}}_{
    (i,j),(h_{j-1},h_j,u_j,d_j)},
    \end{aligned}
    \label{eq:row_transfer}
\end{equation}
where $h_j$ denotes the virtual bond connecting
neighboring tensors within row $i$, with
$h_0=h_{L_y}=1$ for open boundaries.
The multi-indices
\begin{equation}
    \mathbf u=(u_1,\ldots,u_{L_y}),
    \qquad
    \mathbf d=(d_1,\ldots,d_{L_y})
\end{equation}
collect the incoming and outgoing inter-row bonds,
respectively. Thus, $C_i(\bx_{[i]})$ contracts
all virtual bonds within row $i$, while leaving
the bonds connecting it to adjacent rows open.

To construct the row-conditional probabilities,
we introduce two boundary environments.
The top boundary $T_{i-1}(\mathbf u)$ contains
the contraction of all previously sampled rows,
with their physical indices fixed. The bottom
environment $B_i(\mathbf d)$ contains the
contraction of all subsequent rows with their
physical indices summed over.

The bottom environments are constructed using
the fully marginalized row transfer operator
\begin{equation}
    \overline C_i
    =
    \sum_{\bx_{[i]}}C_i(\bx_{[i]}).
\end{equation}
Starting from the trivial boundary $B_{L_x}=1$,
they satisfy the recursion
\begin{equation}
    B_{i-1}(\mathbf u)
    =
    \sum_{\mathbf d}
    [\overline C_i]_{\mathbf d,\mathbf u}
    B_i(\mathbf d).
    \label{eq:bottom_update}
\end{equation}
Since the bottom environments are independent of the
sampled configurations, they can be precomputed and
reused throughout the sampling procedure.

Combining the two boundaries with the transfer operator
of the current row gives the unnormalized conditional weight
\begin{equation}
    \begin{aligned}
    &M_i(\bx_{[i]}\mid\bx_{[<i]})\\
    &=
    \sum_{\mathbf u,\mathbf d}
    T_{i-1}(\mathbf u)
    [C_i(\bx_{[i]})]_{\mathbf d,\mathbf u}
    B_i(\mathbf d).
    \end{aligned}
    \label{eq:row_marginal}
\end{equation}
The corresponding conditional probability is
\begin{equation}
    p_\theta(\bx_{[i]}\mid\bx_{[<i]})
    =
    \frac{M_i(\bx_{[i]}\mid\bx_{[<i]})}
    {\displaystyle\sum_{\bs_{[i]}}
    M_i(\bs_{[i]}\mid\bx_{[<i]})}.
    \label{eq:row_conditional}
\end{equation}

For fixed top and bottom boundaries, the contraction in
Eq.~\eqref{eq:row_marginal} defines an effective
one-dimensional tensor network along row $i$.
We therefore sample the physical indices within each row
sequentially using the 1-norm TT sampling procedure
described in Refs.~\cite{vidal_sample,dolgov_sample,ZZ_TTEDA}.
After sampling row $\bx_{[i]}$, its contribution is
incorporated into the top boundary through
\begin{equation}
    T_i(\mathbf d)
    =
    \sum_{\mathbf u}
    T_{i-1}(\mathbf u)
    [C_i(\bx_{[i]})]_{\mathbf d,\mathbf u},
    \label{eq:top_update}
\end{equation}
starting from $T_0=1$. Repeating this procedure
row by row generates a complete configuration $\bx$.

With exact contractions, the sequential procedure generates
configurations distributed according to $p_\theta(\bx)$.
For a two-dimensional tensor network, however, the boundary bond
dimension generally grows exponentially as successive
rows are incorporated. To make the procedure tractable,
we employ the boundary-MPS contraction
method~\cite{Verstaete_PEPSAlgo,Lubasch_PEPSContactions},
truncating the boundary bond dimension to $\chi$.
For a single-layer network, the dominant computational
cost per site scales as
$\mathcal{O}(\chi^3D^3+\chi^2D^4)$
~\cite{Lubasch_PEPSContactions}.
We use $\chi=3D$ with a singular-value cutoff of $10^{-16}$,
both for sampling and for the environments used in the
update, motivated by previous numerical
studies~\cite{PEPS_Direct_Sampling, Puente_VariFPEPS}.
Although the exact marginal weights
are non-negative, numerical truncation may produce
negative values. These are clipped to zero before
normalization.

At finite $\chi$, approximate contractions can
introduce deviations from the target distribution
$p_\theta(\bx)$.       
Each sampled configuration is nevertheless evaluated
directly using the black-box objective function,
so contraction errors affect the sampling distribution
but not the objective evaluations.

\subsection{Updating Search Distribution}
\label{subsec:Update}

Sampling and evaluating candidate configurations identifies which parts of the search space yield lower objective values. We use this information to update the search distribution so that subsequent samples are more likely to explore these regions. At iteration $t$, we select the $M$ configurations
with the lowest objective values, forming the elite
set $\mathcal E_t=\{\bx^{(r)}\}_{r=1}^{M}$.
We increase their probability by minimizing the
mean negative log-likelihood,
\begin{equation}
    \mathcal J(\theta)
    =
    -\frac{1}{M}\sum_{r=1}^{M}
    \log p_\theta(\bx^{(r)}).
    \label{eq:likelihood}
\end{equation}
Using the 1-norm parametrization in
Eq.~\eqref{eq:onenorm}, this becomes
\begin{equation}
    \mathcal J(\theta)
    =
    \log Z_\theta
    -
    \frac{1}{M}\sum_{r=1}^{M}
    \log W_\theta(\bx^{(r)}).
    \label{eq:nll}
\end{equation}
The first term accounts for the normalization of the
distribution, while the second favors the selected
configurations. In the following, we consider two approaches to minimizing this loss: a gradient-based update and a local sweeping
method that exploits the multilinear structure of tensor networks.

\paragraph{Global gradient descent}
A straightforward approach to updating the search distribution
is to optimize all tensors jointly using gradient descent
on the negative log-likelihood $\mathcal J$. We want to emphasize that this is the gradient with respect to the distribution parameters, not that of the objective function.
For each site tensor $A_k$, a gradient descent step takes
the form
\begin{equation}
    A_k \leftarrow A_k
    -\eta\nabla_{A_k}\mathcal J,
    \label{eq:gradient_update}
\end{equation}
where $\eta$ is the learning rate.
Differentiating Eq.~\eqref{eq:nll} with respect to
the parameters $\theta=\{A_k\}$ gives
\begin{equation}
    \nabla_\theta\mathcal J
    =
    \frac{\nabla_\theta Z_\theta}{Z_\theta}
    -
    \frac{1}{M}\sum_{r=1}^{M}
    \frac{\nabla_\theta W_\theta(\bx^{(r)})}
    {W_\theta(\bx^{(r)})}.
    \label{eq:grad}
\end{equation}
Both terms can be evaluated using boundary-MPS
contractions. For the configuration weights,
the physical indices are fixed to their values
in the elite configurations, whereas for the
normalization they are summed over.

To express the gradients explicitly, we introduce
two local contraction environments.
Let $E^{(r)}_{k,\boldsymbol\alpha}$ denote the
environment of site $k$, obtained by contracting
all other tensors with their physical indices
fixed to the elite configuration $\bx^{(r)}$,
\begin{equation}
    E^{(r)}_{k,\boldsymbol\alpha}
    =
    \vcenter{\hbox{
    \includegraphics[width=.48\linewidth]
    {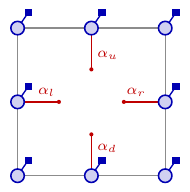}}}.
    \label{eq:data_environment_diagram}
\end{equation}
Similarly, let
$G_{k,\boldsymbol\alpha}
=\partial Z_\theta/\partial A^s_{k,\boldsymbol\alpha}$
denote the normalization environment, obtained
by summing over the physical indices of all
other sites,
\begin{equation}
    G_{k,\boldsymbol\alpha}
    =
    \vcenter{\hbox{
    \includegraphics[width=.48\linewidth]
    {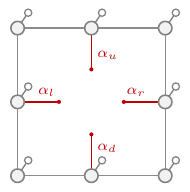}}}.
    \label{eq:normalization_environment_diagram}
\end{equation}

Using these environments, the gradient with
respect to an individual tensor entry becomes
\begin{equation}
    \begin{aligned}
    \frac{\partial\mathcal J}
    {\partial A^s_{k,\boldsymbol\alpha}}
    &=
    \frac{G_{k,\boldsymbol\alpha}}{Z_\theta}\\
    &\quad-
    \frac{1}{M}\sum_{r=1}^{M}
    \frac{
    E^{(r)}_{k,\boldsymbol\alpha}
    \delta_{s,x_k^{(r)}}
    }{
    W_\theta(\bx^{(r)})
    }.
    \end{aligned}
    \label{eq:tensor_gradient}
\end{equation}
To preserve non-negativity during gradient updates,
we parameterize the tensor entries as $A=g(\phi)$,
where $g:\mathbb{R}\to\mathbb{R}_{\geq0}$ is applied
entrywise. The corresponding gradients are obtained
via the chain rule,
\begin{equation}
    \frac{\partial\mathcal J}
    {\partial\phi^s_{k,\boldsymbol\alpha}}
    =
    g'(\phi^s_{k,\boldsymbol\alpha})
    \frac{\partial\mathcal J}
    {\partial A^s_{k,\boldsymbol\alpha}}.
\end{equation}
We perform the optimization using Adam~\cite{kingma_ADAM}.

\paragraph{Multiplicative local sweeps.}

While the joint gradient approach optimizes all tensors simultaneously, the multilinear structure of tensor networks provides
an alternative: When all tensors except one are held fixed, both the configuration weights and the normalization depend linearly on the remaining tensor. We exploit this property to perform alternating optimization, updating one tensor at a time while keeping the others fixed. The resulting update is an instance of the multiplicative Kullback--Leibler (KL) update used in non-negative matrix and tensor factorization~\cite{lee_learning,hood2026nearuniversal}. With exact contraction environments, each multiplicative step is nonincreasing in the KL divergence from the empirical distribution of the selected configurations to the PEPS distribution. Equivalently, their mean log-likelihood is nondecreasing.

Consider updating the tensor $A_k$ at site $k$. Using the contraction environments introduced above, we can express the configuration weights and
normalization as
\begin{align}
    W_\theta(\bx^{(r)})
    &=
    \sum_{\boldsymbol\alpha}
    E^{(r)}_{k,\boldsymbol\alpha}
    A^{x_k^{(r)}}_{k,\boldsymbol\alpha},
    \label{eq:local_weight}\\
    Z_\theta
    &=
    \sum_{s,\boldsymbol\alpha}
    G_{k,\boldsymbol\alpha}
    A^s_{k,\boldsymbol\alpha},
    \label{eq:local_norm}
\end{align}
where $x_k^{(r)}$ corresponds to the value at site $k$ of the sampled configuration $r$.
Using the above equations, the local loss is written as
\begin{equation}
    \begin{aligned}
    \mathcal J_k(A_k)
    &=
    \log\left(
    \sum_{s,\boldsymbol\alpha}
    G_{k,\boldsymbol\alpha}
    A^s_{k,\boldsymbol\alpha}
    \right)\\
    &\quad-
    \frac{1}{M}\sum_{r=1}^{M}
    \log\left(
    \sum_{\boldsymbol\alpha}
    E^{(r)}_{k,\boldsymbol\alpha}
    A^{x_k^{(r)}}_{k,\boldsymbol\alpha}
    \right).
    \end{aligned}
    \label{eq:site}
\end{equation}
In this way, we reduced the global optimization problem to a local one, and the only variational parameter is the corresponding site tensor. Since multiplying $A_k$ by a positive constant rescales
both the configuration weights and the normalization
equally, the probability distribution remains unchanged.
We can therefore fix the scale of $A_k$ by imposing
\begin{equation}
    \sum_{s,\boldsymbol\alpha}
    G_{k,\boldsymbol\alpha}
    A^s_{k,\boldsymbol\alpha}
    =
    Z_0,
    \qquad Z_0>0.
    \label{eq:local_constraint}
\end{equation}
Under this constraint, the normalization term in
Eq.~\eqref{eq:site} becomes constant, and the local
fitting problem reduces to
\begin{equation}
    \begin{aligned}
    \max_{A_k\geq0}\quad&
    \frac{1}{M}\sum_{r=1}^{M}
    \log\left(
    \sum_{\boldsymbol\alpha}
    E^{(r)}_{k,\boldsymbol\alpha}
    A^{x_k^{(r)}}_{k,\boldsymbol\alpha}
    \right)\\
    \text{subject to}\quad&
    \sum_{s,\boldsymbol\alpha}
    G_{k,\boldsymbol\alpha}
    A^s_{k,\boldsymbol\alpha}=Z_0.
    \end{aligned}
    \label{eq:local_optimization}
\end{equation}
The objective is a sum of logarithms of positive
linear functions of $A_k$ and is therefore concave.
Maximizing this objective subject to the linear
normalization constraint and non-negativity of the
tensor entries constitutes a convex optimization
problem. However, this does not imply convexity
of the joint optimization over all tensors, which
remains generally non-convex due to their multiplicative coupling.

Introducing a Lagrange multiplier $\lambda$,
we write
\begin{equation}
    \begin{aligned}
    \mathcal L_k(A_k,\lambda)
    &=
    \frac{1}{M}\sum_{r=1}^{M}
    \log W_\theta(\bx^{(r)})\\
    &\quad-
    \lambda\left(
    \sum_{s,\boldsymbol\alpha}
    G_{k,\boldsymbol\alpha}
    A^s_{k,\boldsymbol\alpha}-Z_0
    \right).
    \end{aligned}
    \label{eq:local_lagrangian}
\end{equation}
Stationarity of the Lagrangian, together with the non-negativity
constraint, gives
\begin{equation}
    \frac{1}{M}\sum_{r=1}^{M}
    \frac{
    E^{(r)}_{k,\boldsymbol\alpha}
    \delta_{s,x_k^{(r)}}
    }{
    W_\theta(\bx^{(r)})
    }
    \leq
    \lambda G_{k,\boldsymbol\alpha},
    \label{eq:local_stationarity}
\end{equation}
with equality wherever
$A^s_{k,\boldsymbol\alpha}>0$.
Multiplying by $A^s_{k,\boldsymbol\alpha}$
and summing over $(s,\boldsymbol\alpha)$
gives $\lambda=1/Z_0$. The resulting fixed-point condition gives
the multiplicative update
\begin{equation}
    A^s_{k,\boldsymbol\alpha}
    \leftarrow\;
    A^s_{k,\boldsymbol\alpha}
    \frac{Z_0}{MG_{k,\boldsymbol\alpha}}\sum_{r=1}^{M}
    \frac{
    E^{(r)}_{k,\boldsymbol\alpha}
    \delta_{s,x_k^{(r)}}
    }{
    W_\theta(\bx^{(r)})
    }.
    \label{eq:em}
\end{equation}
For  non-negative environments and positive
initialization, this update
preserves non-negativity and the local normalization,
while monotonically increasing the local log-likelihood.
The concavity of the local objective ensures that
any solution satisfying the stationarity conditions
is globally optimal for the local problem. With the approximate environments of a PEPS, these guarantees need not hold.

We employ an alternating optimization procedure akin to single-site DMRG and sweep through the lattice. In each local step of the sweep, we update one tensor at a time. Each update is then incorporated into the environments of subsequent sites. The contraction environments are cached and reused throughout the sweep to reduce computational cost. The local fit increases the likelihood of the current elites, but these
configurations are only the best candidates encountered so far. Prematurely concentrating the distribution around them can prevent the search from looking for better regions. To mitigate this, we introduce two tempering mechanisms. First, \emph{uniform smoothing}
mixes the updated tensor with a uniform
tensor $U_k$, whose entries satisfy the same local
normalization constraint,
\begin{equation}
    (U_k)^s_{\boldsymbol\alpha}
    =
    \frac{Z_0}{
    d\sum_{\boldsymbol\alpha'}
    G_{k,\boldsymbol\alpha'}
    }.
    \label{eq:uniform_prior}
\end{equation}
Let $A_k^\ast$ denote the tensor obtained after the
multiplicative iterations. We smooth this tensor as
\begin{equation}
    A_k^\ast
    \leftarrow
    (1-\alpha)A_k^\ast+\alpha U_k,
    \qquad 0\leq\alpha\leq1.
    \label{eq:prior}
\end{equation}
Second, \emph{damping} limits the change in the tensor
during each update. Rather than replacing the old tensor
directly with $A_k^\ast$, we use
\begin{equation}
    A_k^{\mathrm{new}}
    \leftarrow
    (1-\eta)A_{k}^{\mathrm{old}}+\eta A_k^\ast,
    \qquad 0<\eta\leq1.
    \label{eq:lin}
\end{equation}
Here, $\eta=1$ corresponds to a full update toward the
fitted tensor, while smaller values give more
conservative updates. Uniform smoothing limits premature
concentration of the distribution, whereas damping
controls the magnitude of each local update.
We anneal $\alpha$ linearly from its initial value
$\alpha_0$ to zero over the optimization.

To maintain exploration independently of the fitting method,
we introduce \emph{mutation}, which acts directly on sampled
configurations before objective evaluation. Each sampled site
is independently resampled uniformly with probability
$\varepsilon_k$. For a binary site, the probability of an
actual bit flip is therefore $\varepsilon_k/2$. The mutation rate depends on the encoding. For direct encoding,
each variable is resampled with probability $1/V$. For quantics
encoding, we use a scale-dependent rate,
\begin{equation}
    \varepsilon_j
    =
    \frac{1}{V}\,2^{-(R-j)},
    \qquad j=1,\ldots,R,
\end{equation}
where $j=1$ denotes the most significant bit. The expected
number of bit flips per candidate is $1-2^{-R}$,
independent of $V$. In binary encoding, flipping bit $j$
changes the discretized level by $\pm 2^{R-j}$.
The mutation schedule therefore favors smaller changes
while allowing less frequent moves at larger scales.

\begin{figure}[tb]
    \centering
    \includegraphics[width=\linewidth]{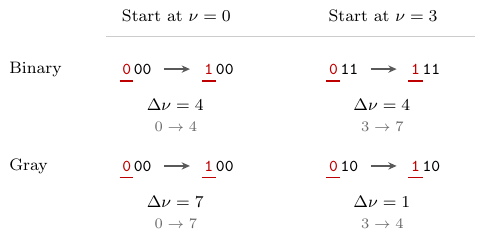}
    \caption{
    Effect of flipping the most significant bit in binary
    and Gray encoding on a three-bit grid. The mutated bit
    is highlighted in red and underlined.
    In binary encoding, the flip always changes the
    numerical level by four. In Gray encoding, the same
    bit flip changes the level by seven from $\nu=0$,
    but only by one from $\nu=3$.
    Binary-space mutation avoids this dependence on
    the starting value by applying bit changes to the
    decoded binary representation.
    }
    \label{fig:mutation_grid}
\end{figure}

However, applying the same schedule directly to bits in
Gray encoding would not preserve these step sizes, since
flipping one bit can change several bits in the decoded
binary representation. As illustrated in
Fig.~\ref{fig:mutation_grid}, flipping the most significant
bit in binary encoding always changes the level by four,
whereas in Gray encoding, the same operation changes
the level by seven from $\nu=0$ but only by one from
$\nu=3$.

To preserve the intended mutation scales, we therefore
perform mutation in binary space. We decode the Gray-coded
variables into binary, apply the mutation schedule, and
convert the result back to Gray code using
\begin{equation}
    \nu_{\mathrm G}
    =
    \nu_{\mathrm B}\oplus(\nu_{\mathrm B}\gg1),
\end{equation}
where $\nu_{\mathrm B}$ and $\nu_{\mathrm G}$ denote the
binary and Gray-code integers, respectively, and $\gg1$
denotes a one-bit right shift. This \emph{binary-space mutation}
preserves the scale-dependent moves of binary encoding while
retaining the Gray-code representation in the PEPS.
Mutated candidates participate in selection, allowing
successful variations to influence subsequent model updates.

Together sampling, mutation, selection, and fitting form the iterative optimization procedure introduced in Fig.~\ref{fig:general_framework}. At each iteration, the selected configurations guide the update of the search distribution, while mutation allows the search to explore configurations beyond those favored by the current model. The procedure is repeated until the budget for the objective evaluation is exhausted, and the best configuration encountered is returned.


\section{Results}
\label{seq:results}
The framework above requires two design choices to be made: how the optimization variables are represented in PEPS, and how the distribution is updated from the selected samples. We first examine these choices on various
continuous functions, where different landscapes allow us to
compare encodings and update rules. The resulting choices are then used for two applications: finding
low-energy configurations of lattice spin glasses and optimizing QAOA
circuits. These applications test the method in rather different
regimes, from thousands of discrete spin variables to a set of
continuous parameters controlling a deep quantum circuit.

Unless stated otherwise, each iteration draws $K=50$ samples and retains
$M=10$ elites. We use damping $\eta=0.75$ and decrease the smoothing
weight linearly from $\alpha_0=0.02$ to zero. The initial PEPS represents
a uniform distribution, and 200 iterations give a budget of $10^4$
objective evaluations. Results are summarized by medians and
25th--75th percentiles over independent runs.

\subsection{Function Optimization}
\label{subsec:functions}
\begin{figure}[tb]
\centering
\includegraphics[width=\columnwidth]{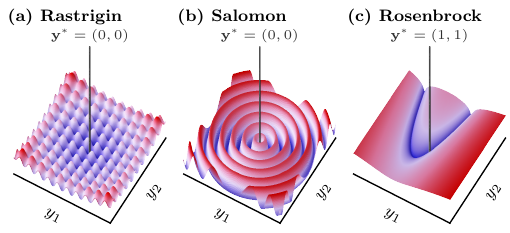}
\caption{Landscapes of three test functions for $V=2$: (a) separable Rastrigin, (b) radial Salomon, and (c) coupled Rosenbrock. Height and color run from low (blue) through purple to high (red). Vertical lines mark the global minimizers $\by^\ast$. }
\label{fig:landscapes}
\end{figure}

\begin{figure}
    \centering
    \includegraphics[width=1\linewidth]{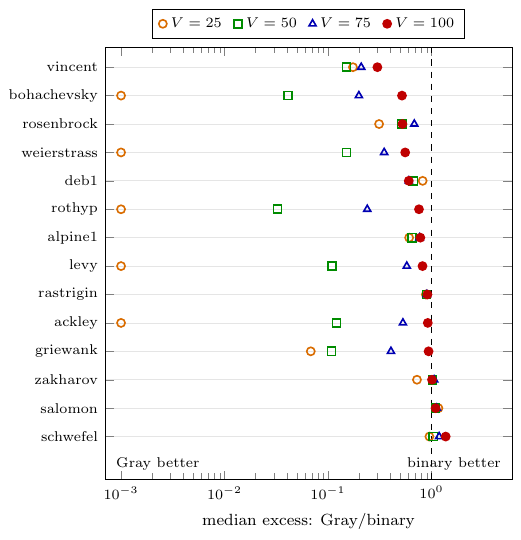}
    \caption{{Gray versus binary code on the 13 functions ($R=6$, $D=2$, $10^4$
evaluations, 50 seeds): ratio of the final median excess $f-f^\ast$ with
Gray code to that with binary code for $V=25$--$100$. Values left
of the dashed line favor Gray code; ratios are floored at $10^{-3}$ where
Gray code finds the exact optimum. Functions are sorted by the ratio at
$V=100$.}}
    \label{fig:graybin_V}
\end{figure}

Continuous test functions provide a controlled setting for studying
how the choices made in Sec.~\ref{seq:theory} affect the search.
We consider 13 functions with separable, radial, and coupled landscapes,
so that the comparisons include different forms of dependence between
variables. Figure~\ref{fig:landscapes} shows one example of each type
for $V=2$. Their definitions, intervals, and minima are given in
Appendix~\ref{app:functions}. We report the excess $f-f^\ast$ relative
to the continuous minimum. The intervals are chosen so that a known
minimizer lies on the discretization grid, except for Schwefel, which
retains its standard domain. Its off-grid minimum gives a finite
discretization floor. The main comparisons use $V=25$,
$50$, $75$, and $100$ variables, $R=6$ bits per variable, $D=2$,
50 seeds per setting, binary-space mutation at finest-bit rate $1/V$,
and $10^4$ evaluations.

\subsubsection{Binary and Gray representations} 
Binary and Gray encodings describe the same discretized
search space but impose different notions of locality on
its representation in the PEPS. To isolate the effect of
this representation, we compare the two encodings using
the same multiplicative sweep and mutation procedure.

Figure~\ref{fig:graybin_V} shows that preserving locality
between neighboring grid points generally improves the
optimization. At $V=50$, Gray encoding achieves a lower
median excess on 10 of the 13 functions, with particularly
pronounced improvements for Griewank and the rotated
hyper-ellipsoid. The advantage extends across different
landscape classes, suggesting that the encoding affects
the ability of the search distribution to resolve promising
regions independently of the particular form of the
objective function.

The difference between the two encodings becomes less
pronounced at $V=100$ under a fixed evaluation budget.
However, increasing the budget to $2\times10^4$ evaluations
restores a clearer separation, with some of the largest
improvements emerging during the later stages of the search
(see Appendix~\ref{app:convergence}). This suggests that
Gray encoding becomes particularly beneficial as the
distribution concentrates and the optimization requires
finer adjustments to individual variables. We therefore adopt Gray encoding for the remaining
benchmarks.

\subsubsection{Multiplicative local sweeps and global gradient updates}

\begin{figure}
    \centering
    \includegraphics[width=1\linewidth]{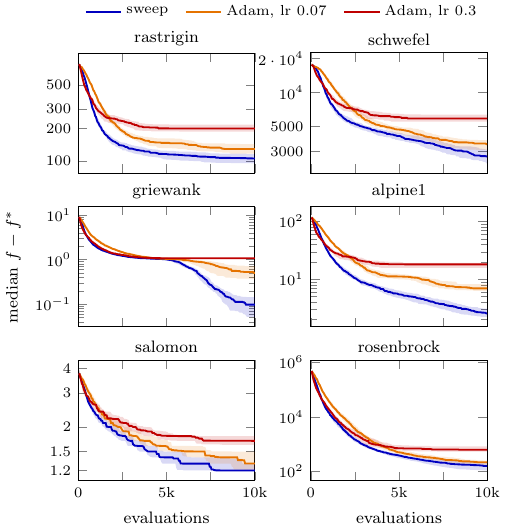}
    \caption{Global gradient updates versus the multiplicative local sweep on six functions ($V=50$, $R=6$, $D=2$,
$10^4$ evaluations, 50 seeds): median excess $f-f^\ast$
versus the number of objective evaluations; shaded regions are the
25th--75th percentiles over seeds.}
    \label{fig:adam_sweep}
\end{figure}

Having selected Gray encoding, we compare the two
approaches to fitting the search distribution introduced
in Sec.~\ref{subsec:Update}: multiplicative local sweeps
and global gradient updates using Adam.

For Adam, we consider three non-negative parameterizations:
square ($g(\phi)=\phi^2$), exponential ($g(\phi)=e^\phi$),
and softplus ($g(\phi)=\log(1+e^\phi)$), each with learning
rates of $0.07$ and $0.3$. The square parameterization
gives the best overall performance among these settings.

As shown in Fig.~\ref{fig:adam_sweep}, the multiplicative
sweep achieves faster convergence and lower final objective
values on all six tested functions, even when the best
Adam setting is selected separately for each function. These results suggest that exploiting the multilinear
structure of the PEPS when fitting the search distribution
leads to more effective optimization of the black-box
objective. This observation is consistent with the
advantages of multiplicative updates over Adam reported
for non-negative tensor factorization
~\cite{hood2026nearuniversal}. We therefore employ multiplicative local sweeps in all
subsequent benchmarks.

\subsubsection{Quantics and direct encoding}

\begin{figure}
    \centering
    \includegraphics[width=1\linewidth]{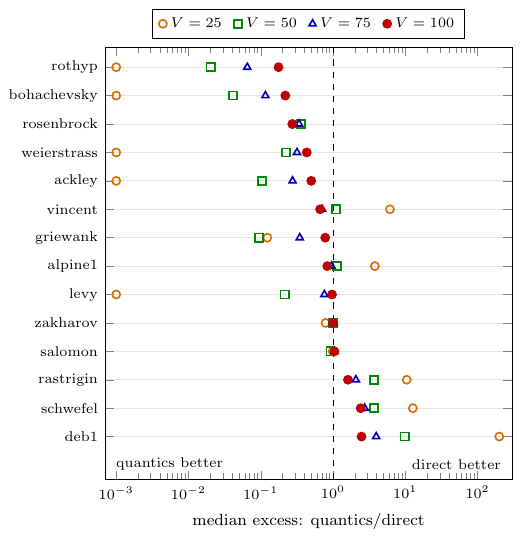}
    \caption{Quantics PEPS versus direct TT on 13 functions ($R=6$, $D=2$ for
both, $10^4$ evaluations, 50 seeds): ratio of the final median excess
$f-f^\ast$ of the quantics encoding  to that of the
direct $d=64$ encoding. Values left of the dashed line favor the
quantics encoding. Ratios are floored at $10^{-3}$ where the quantics
encoding finds the exact optimum and capped at $200$ where the direct
encoding does. Functions are sorted by the ratio at
$V=100$.}
    \label{fig:encoding_V}
\end{figure}
The preceding comparisons employ a quantics representation, in which
each discretized variable is resolved into a sequence of binary sites.
We now assess this choice against a direct encoding, where each variable
is represented by a single site, as in the tensor train (TT) approaches of
Refs.~\cite{PROTES,ZZ_TTEDA}. Figure~\ref{fig:encoding_V} compares
the $V\times6$ quantics PEPS with a direct TT consisting of $V$
sites with local physical dimension $d=64$. Both representations
describe the same 64 discretization levels per variable and use
bond dimension $D=2$, but differ in how these degrees of freedom
are distributed and connected within the tensor network.
Across the benchmark, quantics achieves better overall
performance at all four system sizes.

The relative performance, however, depends on the objective function.
Quantics consistently outperforms direct encoding on the rotated
hyper-ellipsoid, Bohachevsky, Rosenbrock, Weierstrass, Ackley, and
Griewank functions. In contrast, direct encoding performs better
on Rastrigin, Schwefel, and Deb1 at every tested size. These three functions are separable, so their global minima
can be obtained by minimizing each variable independently.
Consequently, correlations between variables are not required
to identify the optimum. Direct encoding may therefore offer
an advantage by representing the probability distribution
over each variable with a single tensor, rather than
distributing it across several binary sites. Separability alone does not
determine the outcome, however, since Weierstrass is also separable but
favors the quantics encoding. Thus, while quantics performs
better overall, direct encoding can be advantageous for particular
objective functions.

\subsubsection{Comparison with Nevergrad}

\begin{figure}
    \centering
    \includegraphics[width=1\linewidth]{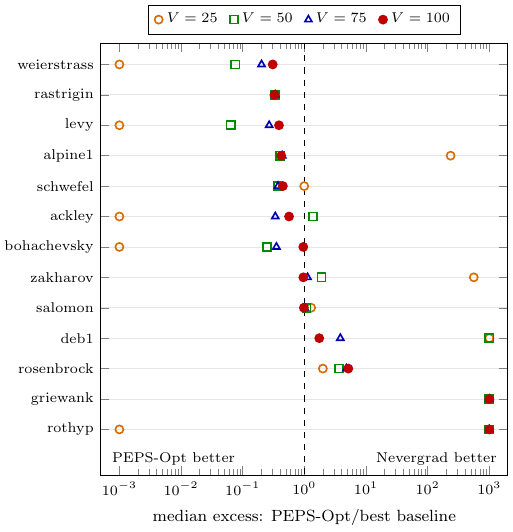}
    \caption{PEPS-Opt $(D=2, R=6)$  versus the strongest Nevergrad optimizer (CMA-ES, DE or PSO) on the 13 functions ($10^4$ evaluations, 50 seeds): ratio of the final median
excess $f-f^\ast$ of the PEPS-Opt to that of the
Nevergrad optimizer, for $V=25$--$100$. Values left of the dashed line favor the PEPS. Ratios are
clipped to $[10^{-3},10^{3}]$: $10^{-3}$ where the PEPS finds the exact optimum, $10^3$ where
CMA-ES reaches machine precision.
The PEPS-Opt searches a discrete grid and the Nevergrad methods search
the continuous interval.
Functions are sorted by the ratio at $V=100$.}
\label{fig:ngfunc_V}
\end{figure}

\begin{figure*}
    \centering
    \includegraphics[width=1\linewidth]{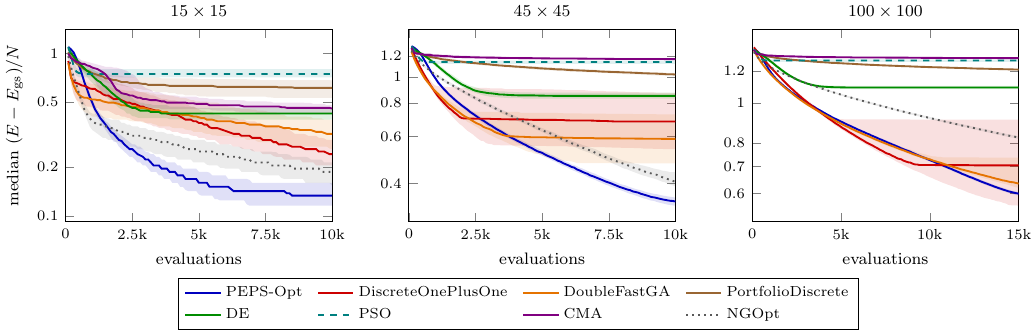}
    \caption{Ising spin glass: median gap per spin
$(E_{\rm best}-E_{\rm gs})/N$ versus the number of objective evaluations
for PEPS-Opt (blue, $D=2$) and seven Nevergrad
optimizers at default settings, 5 instances $\times$ 50 seeds. The budget is
$10^4$ evaluations at $15\times15$ and $45\times45$ and $1.5\times10^4$ at
$100\times100$. The initial smoothing weight is $\alpha_0=0.05$, $0.0167$ and $0.01$ at the three sizes. Shaded
regions are the 25th--75th percentiles.}
    \label{fig:sg_ng}
\end{figure*}

We now compare the resulting PEPS-Opt framework with
CMA-ES, differential evolution (DE), and particle swarm
optimization (PSO), using their default implementations
in Nevergrad~\cite{nevergrad}. The methods are compared
under the same objective evaluation budget, although
the Nevergrad optimizers operate on the continuous
domain while PEPS-Opt searches a discretized grid.

Figure~\ref{fig:ngfunc_V} reveals a marked dependence
of the relative performance on the objective landscape.
PEPS-Opt consistently outperforms the tested baselines
on Rastrigin, Weierstrass, and Levy, and also performs
better on Alpine1 and Schwefel for $V\geq50$. These
functions are characterized by numerous local minima
or rapidly varying objective values, which complicate
the progressive refinement of individual solutions. 

A possible explanation lies in the structure of the
search distribution. While CMA-ES adapts a single
Gaussian distribution, a PEPS can represent multimodal
distributions over candidate configurations. This allows
the search, in principle, to retain information about
several promising regions rather than concentrating
around a single basin. Such flexibility may also help
counter the loss of population diversity that can occur
in DE and PSO. CMA-ES, on the other hand, performs particularly well
on the rotated hyper-ellipsoid, Rosenbrock, and Griewank.
Despite their differing curvature and local structure,
these landscapes possess an underlying basin that
provides useful information for adapting the covariance
and step size of the search distribution. Continuous
optimization also avoids the discretization floor
encountered by PEPS-Opt on functions whose minimizers
do not lie on its grid.

The convergence curves in Appendix~\ref{app:convergence}
provide further insight into these differences. On
functions favoring PEPS-Opt, the competing methods
typically stagnate or improve slowly, while PEPS-Opt
continues to reduce the objective. Conversely, on
functions favoring CMA-ES, much of its advantage emerges
during the final refinement of the solution. These results indicate that the principal strength of
PEPS-Opt lies in navigating rugged and multimodal
landscapes, rather than in the high-precision refinement
of solutions within a smooth basin.

\subsection{Spin Glass}\label{sec:sg}

Lattice spin glasses provide a natural setting in which
to exploit the two dimensional connectivity of a PEPS.
Unlike the continuous benchmarks, where the network
geometry is determined by the encoding, the PEPS here
directly follows the interaction geometry of the
objective function.
Moreover, the Boltzmann distribution
$p_\beta(\boldsymbol{\sigma})\propto e^{-\beta H(\boldsymbol{\sigma})}$
of a nearest-neighbor Ising model admits an exact non-negative
PEPS representation with bond dimension $D=2$, independent of
system size and the presence of frustration.
As the inverse temperature $\beta$ increases, this distribution
becomes increasingly concentrated on low-energy configurations.
Thus, the PEPS ansatz can represent distributions that strongly
favor ground-state configurations, although learning such distributions
through optimization remains a separate challenge that we address here. This approach differs from fitting the PEPS to samples drawn from
a known Boltzmann distribution, which we discuss in
Appendix~\ref{app:gibbs_fit}.

We consider the Edwards--Anderson model
~\cite{Edwards_SpinGlass} on square lattices with open
boundaries,
\begin{equation}
  H(\boldsymbol\sigma)
  =-\sum_{\langle ij\rangle}J_{ij}\sigma_i\sigma_j,
  \qquad \sigma_i,J_{ij}\in\{-1,1\}.
\end{equation}
Each spin is represented by a single PEPS site with
physical dimension $d=2$, using the mapping
$\sigma_i=2x_i-1$. We consider lattice sizes
$L=15$, $45$, and $100$, corresponding to up to
$10^4$ optimization variables. Exact ground-state
energies $E_{\rm gs}$ are obtained using minimum-weight
perfect matching on the dual lattice
~\cite{Barahona_ComputationalComplexitySpinGlass}
and serve exclusively as reference values.

Figure~\ref{fig:sg_ng} compares PEPS-Opt with seven
Nevergrad optimizers operating on the same discrete
search space: DiscreteOnePlusOne, DoubleFastGA,
PortfolioDiscrete, DE, PSO, CMA-ES, and NGOpt.
Performance is measured by the best-found energy gap
per spin, $(E_{\rm best}-E_{\rm gs})/N$, over five
disorder realizations and 50 independent runs per
realization. The evaluation budget is $10^4$ for
$L=15$ and $45$, and $1.5\times10^4$ for $L=100$.

PEPS-Opt achieves the lowest final median energy gap at all three lattice sizes.
It also exhibits smaller run-to-run fluctuations than
its closest competitors, NGOpt at the two smaller
sizes and DoubleFastGA at $L=100$. Several other
optimizers exhibit similarly small fluctuations but
converge to substantially higher energies.

The convergence curves reveal different behavior over the
evaluation budget. The Nevergrad optimizers generally reduce
the energy more rapidly at the beginning of the search,
whereas PEPS-Opt improves more gradually and continues to
find lower-energy configurations.
This distinction becomes particularly apparent at $L=100$,
where PEPS-Opt overtakes the leading Nevergrad methods only
late in the search and is still improving at the end of the
evaluation budget. The remaining gap to the exact ground-state
energy is nevertheless substantial, indicating that these
results demonstrate improved search performance within the
available budget rather than convergence to the ground state.

\begin{figure*}
    \centering
    \includegraphics[width=1\linewidth]{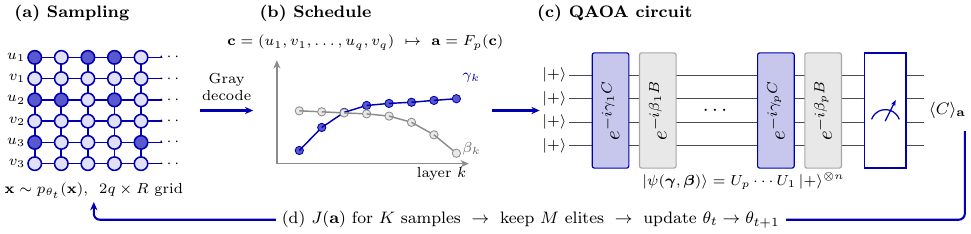}
    \caption{QAOA parameter search with PEPS-Opt using the Fourier encoding. (a) A PEPS on a
$2q\times R$ grid samples bitstrings $\bx$; neighboring rows hold the quantics
bits of the coefficient pairs $(u_i,v_i)$. (b) Gray decoding gives the Fourier
coefficients $\mathbf c$, and the expansion $F_p$ of Eq.~\eqref{eq:fourier}
gives the angles $\gamma_k$ and $\beta_k$ of all $p$ layers. (c) The QAOA
circuit, with $B=\sum_jX_j$, is simulated to obtain $\langle C\rangle_{\mathbf a}$
and the cost $J(\mathbf a)$ of Eq.~\eqref{eq:qaoa_cost}. (d) The $M$
lowest-cost samples of the batch update the PEPS, and the loop repeats. Only
the expansion $F_p$ depends on the depth $p$.}
    \label{fig:qaoa_loop}
\end{figure*}

\begin{figure}
    \centering
    \includegraphics[width=1\linewidth]{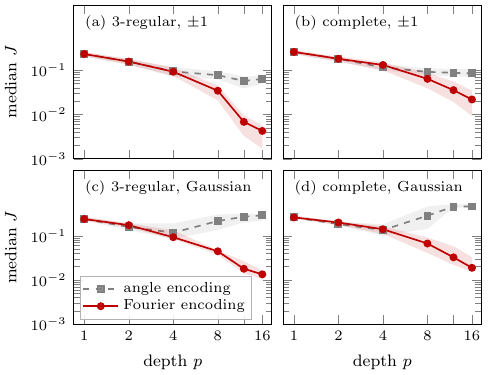}
    \caption{Angle versus Fourier encoding of the QAOA parameters in PEPS-Opt, for MaxCut on $n=12$ qubits:
median normalized cost $J$ at $10^4$ evaluations for $p=1,2,4,8,12,16$ (10 instances
$\times$ 50 seeds per point). Shaded regions are the 25th--75th percentiles.
Both encodings use $D=2$, $R=8$ Gray bits per variable. The angle encoding represents the
$V=2p$ layer angles; the Fourier encoding represents $V=2q$ coefficients with
$q=3$ and expands them through Eq.~\eqref{eq:fourier}. }
    \label{fig:qaoa_encoding}
\end{figure}

\subsection{Quantum Approximate Optimization Algorithm}

Variational quantum algorithms provide a different setting
for black-box optimization, where objective evaluations
involve executing parameterized quantum circuits and
estimating expectation values. We investigate PEPS-Opt
in this setting using the quantum approximate optimization
algorithm (QAOA), focusing on two challenges: the increasing
number of variational parameters with circuit depth and
the stochasticity of objective evaluations arising from
finite measurement statistics.

We consider QAOA for weighted MaxCut~\cite{Farhi_QAOA},
with the circuit state
\begin{align}
  \ket{\psi(\boldsymbol\gamma,\boldsymbol\beta)}
  &=U_p\cdots U_1\ket{+}^{\otimes n},\\
  U_k&=e^{-i\beta_k\sum_jX_j}e^{-i\gamma_k C},
\end{align}
where the cost operator is
$C=\sum_{\langle uv\rangle}w_{uv}(1-Z_uZ_v)/2$.
The objective is to minimize the normalized cost
\begin{equation}
  J(\mathbf a)=\frac{C_{\max}-\langle C\rangle_{\mathbf a}}
  {C_{\max}-C_{\min}}\in[0,1],
  \label{eq:qaoa_cost}
\end{equation}
where $\mathbf a$ denotes the circuit parameters,
$\langle C\rangle_{\mathbf a}$ is the expectation value in
the QAOA state, and $C_{\max}$ and $C_{\min}$ are the
extremal cut values obtained by enumeration.

The benchmarks use $n=12$ qubits and four graph classes:
random 3-regular and complete graphs, each with either
$\pm1$ or Gaussian $\mathcal N(0,1)$ edge weights.
The two 3-regular classes share the same graph topology.
Circuit depths $p=8,16,32,64,128$ are considered, with
ten instances and 50 independent runs per instance,
each limited to $10^4$ objective evaluations.

\subsubsection{Angle and Fourier representations}

\begin{figure*}
    \centering
    \includegraphics[width=1\linewidth]{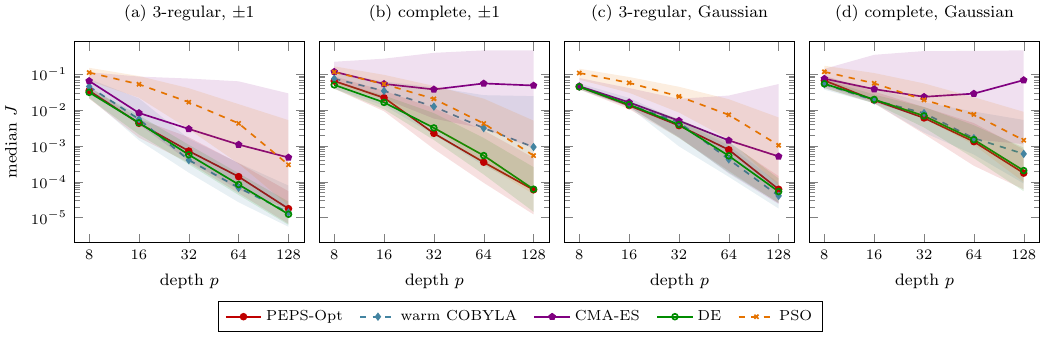}
    \caption{QAOA MaxCut on $n=12$ qubits: median normalized cost $J$ at a nominal
budget of $10^4$ evaluations versus depth $p$ (10 instances $\times$ 50 seeds
per point). Shaded regions are the 25th--75th percentiles. PEPS-Opt,
warm COBYLA, CMA-ES, DE, and PSO search the same $2q=6$ Fourier coefficients
($q=3$) with the same schedule expansion. PEPS-Opt uses $D=2$, and
eight-bit Gray encoding. CMA-ES starts from
a uniform random point; DE and PSO use uniform initial populations. PEPS-Opt starts near a uniform
search distribution. These methods are initialized independently at each depth.
In contrast, warm-started COBYLA follows a depth-doubling
schedule, using the optimized Fourier coefficients from
the preceding depth as its initial guess. COBYLA also
allows coefficients outside the prescribed bounds,
whereas the other methods share the same bounded search
domain, which PEPS-Opt discretizes.}
    \label{fig:qaoa_fourier}
\end{figure*}

For a QAOA circuit of depth $p$, the variational
parameters are given by
\begin{equation}
  \mathbf a=(\gamma_1,\beta_1,\ldots,\gamma_p,\beta_p).
  \label{eq:angle_vector}
\end{equation}
We represent these $2p$ angles on a $2p\times R$
quantics PEPS, with neighboring rows corresponding to
the cost and mixing angles of each layer. We refer to
this representation as the \emph{angle encoding}.

Optimizing each angle independently becomes increasingly
challenging as the circuit depth grows.
Ref.~\cite{Zhou_QAOA} observed that optimized QAOA
schedules often vary smoothly across circuit layers
and can be described using a small number of Fourier
components. Following this approach, we introduce
\begin{equation}
  \mathbf c=(u_1,v_1,\ldots,u_q,v_q),\qquad
  \mathbf a=F_p(\mathbf c),
  \label{eq:fourier_vector}
\end{equation}
where
\begin{equation}
  \gamma_k=\sum_{i=1}^{q}u_i\sin\phi_{ik},
  \qquad
  \beta_k=\sum_{i=1}^{q}v_i\cos\phi_{ik},
  \label{eq:fourier}
\end{equation}
with $\phi_{ik}=(i-\frac12)(k-\frac12)\pi/p$ and
$k=1,\ldots,p$.

We refer to this representation as the \emph{Fourier
encoding}. Unlike the local optimization approach
of Ref.~\cite{Zhou_QAOA}, we learn a joint probability
distribution over the Fourier coefficients.
Related tensor-train approaches to quantum control
represent each coefficient by a single site
~\cite{ZZ_TTEDA}. Here, we instead resolve the
coefficients into quantics bits on a $2q\times R$
PEPS, following the layout in Fig.~\ref{fig:encoding}(b).
Figure~\ref{fig:qaoa_loop} illustrates the optimization
loop. PEPS-Opt evaluates $J(\mathbf a)$ in the angle
encoding and $J(F_p(\mathbf c))$ in the Fourier encoding.
The sampling and fitting procedures are identical
for both representations.

At fixed $q$, the number of Fourier coefficients is
independent of circuit depth. For $q<p$, the expansion
restricts the search to a lower-dimensional space of
smooth schedules, whereas for $q\geq p$ the representation
can be redundant. We fix $q=3$ with coefficient bounds
$|u_i|\leq2.5/i$ and $|v_i|\leq1.2/i$. For the angle
encoding, we use $\gamma_k\in[0,2\pi)$ and
$\beta_k\in[0,\pi/2)$, where the latter exploits the
global spin-flip symmetry of the circuit.
Both representations use $D=2$ and $R=8$ Gray bits
per variable. Appendix~\ref{app:qaoa_graybin} further
shows that Gray encoding outperforms binary encoding
across all tested Fourier settings.

Figure~\ref{fig:qaoa_encoding} compares the two
representations at depths $p=1,2,4,8,12,16$.
Their relative performance varies at smaller depths,
but the Fourier encoding develops a clear advantage
as $p$ increases. At $p=8$ and $16$, it achieves
lower median costs across all four graph classes,
with particularly large improvements at $p=16$.

In the angle encoding, the number of independently
optimized angles grows with circuit depth, making the
search increasingly high-dimensional. The Fourier
representation instead describes the angle schedules
using a fixed number of coefficients, reducing the
search dimension while favoring smooth variations
across circuit layers. Both effects may contribute
to its improved performance at larger depths.

\begin{figure*}
     \centering
     \includegraphics[width=1\linewidth]{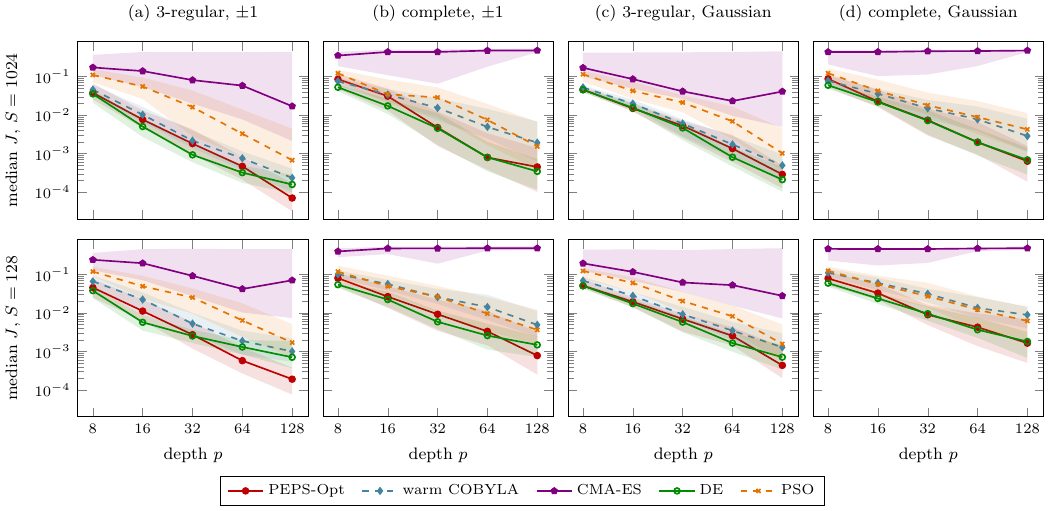}
     \caption{QAOA MaxCut on $n=12$ qubits with shot noise: median normalized cost $J$ versus depth $p$
when every evaluation is estimated from $S=1024$ (top) or $S=128$ (bottom) measurement shots of the QAOA
state (10 instances $\times$ 10 seeds per point). Shaded regions are the 25th--75th percentiles. All methods
search the same $2q=6$ Fourier coefficients with the same initializations as in Fig.~\ref{fig:qaoa_fourier}
and a budget of $10^4$ evaluations.
$J$ is evaluated exactly at the schedule each method returns.}
     \label{fig:qaoa_noise}
 \end{figure*}

\subsubsection{Performance across circuit depths}

We now compare PEPS-Opt with CMA-ES, differential
evolution (DE), particle swarm optimization (PSO),
and warm-started COBYLA. All methods optimize the
same six Fourier coefficients at circuit depths
$p=8,16,32,64,128$.

CMA-ES starts from a uniformly sampled point, while
DE and PSO start from uniform populations.
PEPS-Opt starts from a nearly uniform search
distribution. These methods are initialized
independently at each depth and share the same
coefficient bounds, with PEPS-Opt operating on
their discretized representation.

For COBYLA, we progressively double the circuit depth
and use the optimized coefficients from the preceding
depth as the initial guess. The nominal budget of
$10^4$ evaluations includes all preceding optimization
stages. COBYLA is also allowed to explore coefficients
outside the prescribed bounds. Although this is not
a like-for-like comparison, we include it as a reference
for commonly used QAOA optimization strategies
~\cite{Zhou_QAOA,Lee_QAOA,Campos_TrainingQAOA}.

Figure~\ref{fig:qaoa_fourier} shows that the costs
obtained with PEPS-Opt decrease with circuit depth
across all four graph classes. The median costs reach
$2\times10^{-5}$--$2\times10^{-4}$ at $p=128$.

The relative performance of PEPS-Opt and warm-started
COBYLA depends on the graph class. On 3-regular graphs,
COBYLA achieves slightly lower costs at larger depths,
with differences of at most about a factor of two.
On complete graphs, PEPS-Opt performs better at every
tested depth for $\pm1$ weights and from $p=32$ onward
for Gaussian weights. At $p=128$, its median costs
are lower by factors of approximately 16 and 3,
respectively.

Among the independently initialized optimizers, DE
is the closest competitor to PEPS-Opt and performs
better on $\pm1$-weighted 3-regular graphs at the
largest depths. PEPS-Opt generally outperforms
CMA-ES, except on Gaussian-weighted 3-regular graphs
at smaller depths, and consistently achieves lower
median costs than PSO.

Since all methods use the same Fourier parameterization,
the comparison distinguishes the performance of the
optimizers from the benefits of the
parameterization. In particular, PEPS-Opt achieves lower costs
than warm-started COBYLA on complete graphs at large
depths, without requiring solutions from shallower
circuits.

\subsubsection{Finite-shot objective evaluations}

The preceding benchmarks use exact expectation values.
On quantum hardware, however, these values must be
estimated from a finite number of measurement shots.
We therefore repeat the Fourier comparison using
$S=1024$ and $128$ shots per objective evaluation.
We consider ten instances and ten independent runs
per instance and depth, with all other optimization
settings unchanged.

Under shot noise, the lowest observed cost may
substantially underestimate the true cost. We therefore
evaluate the exact cost of the schedule returned by
each optimizer. This information is not available
during the search.
Rather than returning the configuration with the
lowest observed cost, PEPS-Opt returns the most
probable of 1000 configurations sampled from the
final search distribution. Warm-started COBYLA returns
the configuration with the lowest estimated cost,
while CMA-ES, DE, and PSO return their respective
Nevergrad recommendations.

Figure~\ref{fig:qaoa_noise} shows that PEPS-Opt and
DE perform best under finite-shot noise. For both
shot budgets, they achieve the two lowest median
costs across all four graph classes at $p\geq32$.
With 128 shots, PEPS-Opt achieves the lowest median
cost at $p=128$ in every graph class.

Shot noise has a particularly strong effect on
warm-started COBYLA. For $\pm1$-weighted 3-regular
graphs at $p=128$, its median cost increases by
nearly two orders of magnitude when using 128 shots,
compared with approximately one order of magnitude
for PEPS-Opt. The advantage of COBYLA on 3-regular
graphs observed with exact evaluations thus disappears
under shot noise. PEPS-Opt achieves lower median
costs than COBYLA across all four graph classes
at $p\geq32$.

\section{Conclusion}
\label{seq:conc}
In this work, we have developed a gradient-free probabilistic optimization framework where a non-negative PEPS parameterizes an adaptive search distribution over candidate solutions. The non-negative parametrization allows normalization and marginalization to be performed through single-layer tensor network contractions. This reduces the computational overhead compared to the Born-machine formulation. We have extended the direct sampling procedure to this parametrization and adapted a DMRG-like local sweeping algorithm that exploits the multilinear
structure of the PEPS and the convexity of the resulting local optimization problems. Furthermore, we have introduced a quantics PEPS representation with Gray-code encoding which provides a multiscale representation of discretized continuous optimization variables while preserving locality between neighboring
discretization levels.

Our numerical benchmarks demonstrate that both the representation of the optimization variables and the procedure used to update the search distribution play important roles in the performance of the optimizer. In the case of continuous test functions, Gray-code encoding generally improves upon standard binary encoding. The local sweeping algorithm consistently outperforms the global gradient updates considered in this work. Compared to established gradient-free baselines used in this work, PEPS-Opt is particularly competitive on rugged landscapes with many separated local minima. For two-dimensional Ising spin glasses, PEPS-Opt achieves lower median energy gaps than the considered
baselines across all investigated lattice sizes. The QAOA benchmarks show that combining PEPS-Opt with a Fourier parametrization of the circuit angles
provides an effective optimization strategy at large circuit depths. Our method also exhibits robustness against finite-shot measurement noise, highlighting its potential applicability to variational quantum algorithms.

Several directions remain open for future investigations.
From a methodological perspective, adaptive discretization
strategies where the resolution of individual optimization
variables is dynamically adjusted during the search can
provide a more efficient allocation of computational resources.
In particular, the quantics representation naturally lends
itself to such adaptive multiscale optimization schemes.
Another promising direction is to combine the probabilistic
search with local optimization procedures  \cite{sozykin2025highdimensionaloptimizationlowrank, sv5r-mryz}. Complementing PEPS-Opt with such local refinement could lead to a more effective exploitation of promising regions of the objective landscape while retaining the global exploration capabilities of the probabilistic approach. From a more fundamental perspective, it would be interesting to investigate the relationship between the structure of the objective landscape and the statistical properties of the learned search distribution. In particular, understanding
how the geometry of the landscape, the organization of its
local minima, and the interactions between optimization
variables manifest themselves in the correlation structure
of the search distribution could provide insights into the
conditions under which tensor network representations are
particularly effective. From an application point of view, the robustness observed under finite-shot measurement noise motivates further investigations of PEPS-Opt in optimization problems involving noisy, uncertain, or experimentally obtained objective evaluations. In particular, its application to quantum optimal control and hardware-based variational quantum algorithms could be of interest. 

Beyond optimization, it would also be interesting to explore
the capabilities of the non-negative PEPS parametrization
introduced in this work for generative learning tasks.
The single-layer formulation, together with the direct
sampling and local likelihood-based optimization procedures,
provides an alternative to PEPS-based Born machines \cite{Vieijra_PEPS_Generative}.
Investigating the trade-off between the representational
capacity and computational efficiency of these two
parametrizations in generative modeling constitutes
another promising direction for future research.

\begin{acknowledgments}

A.D. acknowledges funding under Horizon Europe programme HORIZON-CL4-2023-DIGITAL-EMERGING-01-CNECT via the project 101135699  (SPINUS). Parts of the code make use of the ITensor library~\cite{Fishman_ITensor}.
LLMs were used as coding and writing assistants. The authors directed their use, reviewed and verified all outputs, and take responsibility for the resulting code and text.

\end{acknowledgments}

\appendix

\section{Test functions}\label{app:functions}
To specify the continuous benchmarks and the grids, we collect the function definitions and domains used in
Sec.~\ref{subsec:functions}. The functions act on
$\by\in\mathbb{R}^V$, with a common interval
$y_i\in[\ell,\ell+w)$ for all coordinates. Both the direct and quantics
encodings use the grid $y=\ell+wk/2^R$, $k=0,\dots,2^R-1$.
Except for Schwefel, the interval is centered on a known
minimizer, $\ell=y^\ast-w/2$, which is then represented exactly at
$k=2^{R-1}$. We report the excess $f(\by)-f^\ast$ relative to the
continuous minimum.
\allowdisplaybreaks
\begin{gather*}
\text{rastrigin: } f=10V+\sum_{i=1}^{V}\bigl[y_i^2-10\cos(2\pi y_i)\bigr],\\
\text{griewank: } f=\frac{1}{4000}\sum_{i=1}^{V}y_i^2-\prod_{i=1}^{V}\cos\frac{y_i}{\sqrt i}+1,\\
\text{schwefel: } f=418.9829\,V-\sum_{i=1}^{V}y_i\sin\sqrt{|y_i|},\\
\text{alpine1: } f=\sum_{i=1}^{V}\bigl|y_i\sin y_i+0.1\,y_i\bigr|,\\
\text{deb1: } f=-\frac1V\sum_{i=1}^{V}\sin^6(5\pi y_i),\\
\text{levy: } f=\sin^2(\pi w_1)+(w_V-1)^2\bigl[1+\sin^2(2\pi w_V)\bigr]\\
\qquad+\sum_{i=1}^{V-1}(w_i-1)^2\bigl[1+10\sin^2(\pi w_i+1)\bigr],\\
w_i=1+\tfrac{y_i-1}{4},\\
\text{salomon: } f=1-\cos(2\pi\lVert\by\rVert)+0.1\,\lVert\by\rVert,\\
\text{zakharov: } f=\sum_{i=1}^{V}y_i^2+t^2+t^4,\quad t=\sum_{i=1}^{V}\tfrac{i}{2}\,y_i,\\
\text{rotated hyper-ellipsoid: } f=\sum_{i=1}^{V}\sum_{j=1}^{i}y_j^2,\\
\text{rosenbrock: } f=\sum_{i=1}^{V-1}\bigl[100(y_{i+1}-y_i^2)^2+(1-y_i)^2\bigr],\\
\text{ackley: } f=-20\,e^{-0.2\sqrt{\overline{y^2}}}-e^{\overline{\cos(2\pi y)}}+20+e,\\
\text{bohachevsky: } f=\sum_{i=1}^{V-1}\bigl[y_i^2+2y_{i+1}^2-0.3\cos(3\pi y_i)\\
\qquad-0.4\cos(4\pi y_{i+1})+0.7\bigr],\\
\text{weierstrass: } f=\sum_{i=1}^{V}\sum_{k=0}^{20}a^k\cos\bigl(2\pi b^k(y_i+\tfrac12)\bigr)\\
\qquad-V\sum_{k=0}^{20}a^k\cos(\pi b^k),\quad a=\tfrac12,\ b=3.
\end{gather*}
Table~\ref{tab:functions} lists the search intervals, minimizing
coordinates, and continuous minima. The minimizing coordinates of Deb1 are
$y^\ast=\pm0.1,\pm0.3,\dots,\pm0.9$.

\begin{table}[htbp]
\caption{Interval $[\ell,\ell+w)$ of every coordinate, minimizer $y^\ast$
per coordinate and minimum $f^\ast$ of the test functions.}
\label{tab:functions}
\begin{ruledtabular}
\begin{tabular}{lccc}
function & interval & $y^\ast$ & $f^\ast$ \\
\hline
rastrigin & $[-5.12,5.12)$ & $0$ & $0$ \\
griewank & $[-51.2,51.2)$ & $0$ & $0$ \\
schwefel & $[-500,500)$ & $420.9687$ & $0$ \\
alpine1 & $[-10.24,10.24)$ & $0$ & $0$ \\
deb1 & $[-1,1)$ & 10 minimizers & $-1$ \\
levy & $[-9,11)$ & $1$ & $0$ \\
salomon & $[-10.24,10.24)$ & $0$ & $0$ \\
zakharov & $[-5.12,5.12)$ & $0$ & $0$ \\
rot.\ hyper-ellipsoid & $[-5.12,5.12)$ & $0$ & $0$ \\
rosenbrock & $[-4.12,6.12)$ & $1$ & $0$ \\
ackley & $[-5.12,5.12)$ & $0$ & $0$ \\
bohachevsky & $[-10.24,10.24)$ & $0$ & $0$ \\
weierstrass & $[-0.512,0.512)$ & $0$ & $0$ \\
\end{tabular}
\end{ruledtabular}
\end{table}

\begin{figure}[H]
    \centering
    \includegraphics[width=1\linewidth]{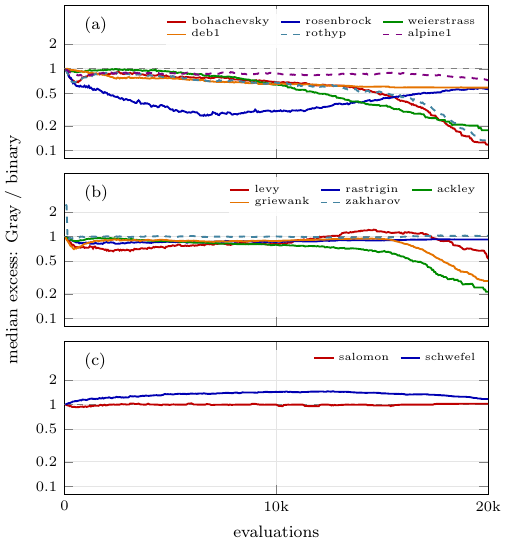}
    \caption{Gray versus binary code at $V=100$ ($R=6$, $D=2$, $2\times10^4$ evaluations, 50 seeds): ratio of the median best-so-far excess with Gray code to that with binary code, versus the number of objective evaluations. Functions are grouped by the outcome of the $10^4$-evaluation runs at $V=100$ (Fig.~\ref{fig:graybin_V}): (a) Gray code ahead, (b) close to a tie, and (c) binary code ahead. Values below 1 favor Gray code.}
    \label{fig:graybin_ratio_V100_20k}
\end{figure}

\begin{figure*}[t]
    \centering
    \includegraphics[width=1\linewidth]{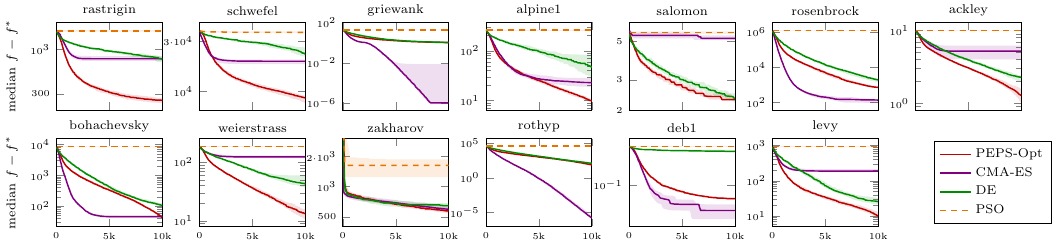}
\caption{PEPS-Opt versus Nevergrad optimizers on the 13 test functions ($V=100$, $10^4$ evaluations,
50 seeds): median best-so-far excess $f-f^\ast$ versus the number of objective evaluations; shaded
regions are the 25th--75th percentiles over seeds. PEPS-Opt searches the $2^6$-level grid. CMA-ES, DE and
PSO (default settings) search the continuous box, CMA-ES from a uniform random point and DE and PSO
from a uniform initial population. Curves are floored at $10^{-6}$.}
\label{fig:ngfunc_conv}
\end{figure*}

\section{Additional convergence comparisons}\label{app:convergence}

The final costs in the main text summarize the outcome of each search,
but do not show when the differences between methods emerge.
To complement those comparisons, Fig.~\ref{fig:graybin_ratio_V100_20k} shows
the binary-versus-Gray comparison at $V=100$ with $2\times10^4$ evaluations
as the ratio of the median excesses, with the functions grouped by the
outcome at $10^4$ evaluations. The largest
differences often emerge late in the run, as the proposal distribution
concentrates and finer-scale changes become important.
Figure~\ref{fig:ngfunc_conv} compares PEPS-Opt with the continuous
Nevergrad baselines at $V=100$. It supplements the final-excess
ratios in Fig.~\ref{fig:ngfunc_V} and shows their differing initial
and late-stage convergence behavior.

\begin{figure}
    \centering
    \includegraphics[width=1\linewidth]{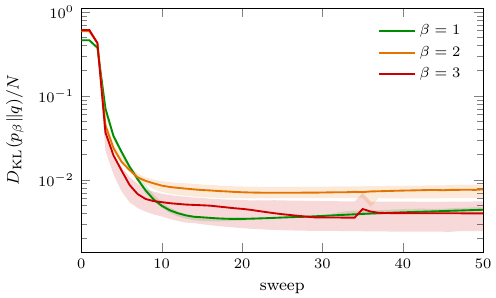}
    \caption{Fitting a $D=2$ PEPS to 2000 samples of the Boltzmann distribution
$p_\beta$ of the $15\times15$ spin glasses, starting from the uniform
distribution: Kullback--Leibler divergence per spin on held-out samples
versus the number of sweeps. Lines are the median over the five instances,
shaded regions the full range.}
    \label{fig:gibbs_fit}
\end{figure}
\section{Fitting the Boltzmann distribution}\label{app:gibbs_fit}

Section~\ref{sec:sg} notes that the Boltzmann distribution
$p_\beta(\boldsymbol\sigma)\propto e^{-\beta H(\boldsymbol\sigma)}$ of a
nearest-neighbor Ising model is an exact non-negative PEPS with $D=2$.
Here we ask whether the multiplicative local sweep, used inside PEPS-Opt,
recovers this distribution from samples when it starts from the
uniform distribution. For each of the five $15\times15$ instances of
Sec.~\ref{sec:sg}, we draw 2000 configurations from the exact
Boltzmann PEPS and fit a $D=2$ PEPS to them with 50 sweeps
and $\beta\in\{1,2,3\}$. We measure the Kullback--Leibler
divergence per spin between $p_\beta$ and the fitted distribution on
2000 held-out samples, using the $\log p_\beta$.

Figure~\ref{fig:gibbs_fit} shows that the sweep recovers $p_\beta$ at
all three temperatures. Within ten sweeps, the divergence falls from
$0.46$--$0.61$ per spin at the uniform start to below $0.01$, and after
50 sweeps it lies between $0.002$ and $0.009$ per spin. The
fitted distributions reproduce the mean energy per spin to within $1.4\%$. 

\section{Binary and Gray code for QAOA}\label{app:qaoa_graybin}

\begin{figure}
    \centering
    \includegraphics[width=1\linewidth]{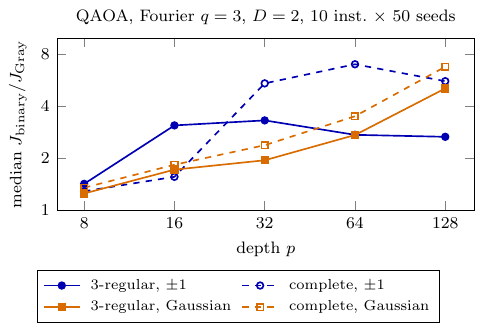}
    \caption{Gray versus binary code for QAOA: ratio of the median normalized cost with binary
code to that with Gray code, Fourier-quantics encoding with $q=3$ and $D=2$
on the depth grid of Fig.~\ref{fig:qaoa_fourier}. Gray code performs better for every
class and depth.}
    \label{fig:graybin}
\end{figure}

The Fourier construction changes which variables are optimized, but
each coefficient must still be encoded in the PEPS. We therefore return
to the binary--Gray comparison to examine whether the encoding effect
seen for the test functions also appears in this schedule subspace.
Figure~\ref{fig:graybin} compares the two codes for the Fourier encoding
under matched settings. Gray code performs
better in all tested settings.
At $p=128$, binary-code median costs are $2.7$--$6.8$ times higher.
The encoding of the Fourier coefficients therefore has a marked
effect on the outcome even within the same schedule subspace.


  

\bibliography{references} 
\end{document}